\pdfoutput=1

\documentclass[11pt]{article}

\usepackage{acl}

\usepackage{times}
\usepackage{latexsym}
\usepackage{booktabs}
\usepackage{graphicx}
\usepackage{comment}
\usepackage{enumitem}
\usepackage{subcaption}
\usepackage{multirow}

\usepackage[T1]{fontenc}

\usepackage[utf8]{inputenc}

\usepackage{microtype}

\usepackage{inconsolata}

\usepackage{graphicx}

\usepackage{quoting}

\title{Missing the Margins: A Systematic Literature Review on\\the Demographic Representativeness of LLMs}

\author{
    Indira Sen\textsuperscript{1}, Marlene Lutz\textsuperscript{1}, Elisa Rogers\textsuperscript{1}, David Garcia\textsuperscript{2}, and {\bf Markus Strohmaier\textsuperscript{1,3}}\\
    \textsuperscript{1}University of Mannheim, 
    \textsuperscript{2}University of Konstanz,
    \textsuperscript{3}GESIS - Leibniz Institute for the Social Sciences\\
    \small
    \texttt{indira.sen@uni-mannheim.de}, \texttt{marlene.lutz@uni-mannheim.de}, \texttt{elisa.marie-rogers@students.uni-mannheim.de},\\
    \small
    \texttt{david.garcia@uni-konstanz.de}, \texttt{markus.strohmaier@uni-mannheim.de}
}

\begin{document}
\maketitle
\begin{abstract}

Many applications of Large Language Models (LLMs) require them to either simulate people or offer personalized functionality, making the demographic representativeness of LLMs crucial for equitable utility. At the same time, we know little about the extent to which these models actually reflect the demographic attributes and behaviors of certain groups or populations, with conflicting findings in empirical research. To shed light on this debate, we review 211 papers on the demographic representativeness of LLMs. We find that while 29\% of the studies report positive conclusions on the representativeness of LLMs, 30\% of these do not evaluate LLMs across multiple demographic categories or within demographic subcategories. Another 35\% and 47\% of the papers concluding positively fail to specify these subcategories altogether for gender and race, respectively. Of the articles that do report subcategories, fewer than half include marginalized groups in their study. \textcolor{black}{Finally, more than a third of the papers do not define the target population to whom their findings apply; of those that do define it either implicitly or explicitly, a large majority study only the U.S. Taken together, our findings} suggest an inflated perception of LLM representativeness in the broader community. We recommend more precise evaluation methods and comprehensive documentation of demographic attributes to ensure the responsible use of LLMs for social applications. Our annotated list of papers and analysis code is publicly available.\footnote{\url{https://github.com/Indiiigo/LLM_rep_review}}

\end{abstract}

\section{Introduction}


In addition to their applications as general assistive technology, an emerging use case of LLMs in the (computational) social sciences is the simulation of human behavior, to replicate or augment existing social data like survey responses~\cite{argyle2023out}, behavioral experiments~\cite{Hewitt2024} or social network traces~\cite{chang2024llms}. For LLMs to be an effective tool in both assisting diverse human populations and simulating their behavior, LLMs would need to be representative, i.e., \textit{their behavior would need to validly reflect the underlying target population.} For example, if providing personalized healthcare or educational recommendations, the LLM should be equally assistive to multiple groups of people, and not display lack of background knowledge for certain groups. Similarly, if an LLM is used in social simulations, then it should also be equally effective at emulating the behavior of different groups of people.

\begin{figure*}[t!]
    \centering
    \includegraphics[width=0.99\linewidth]{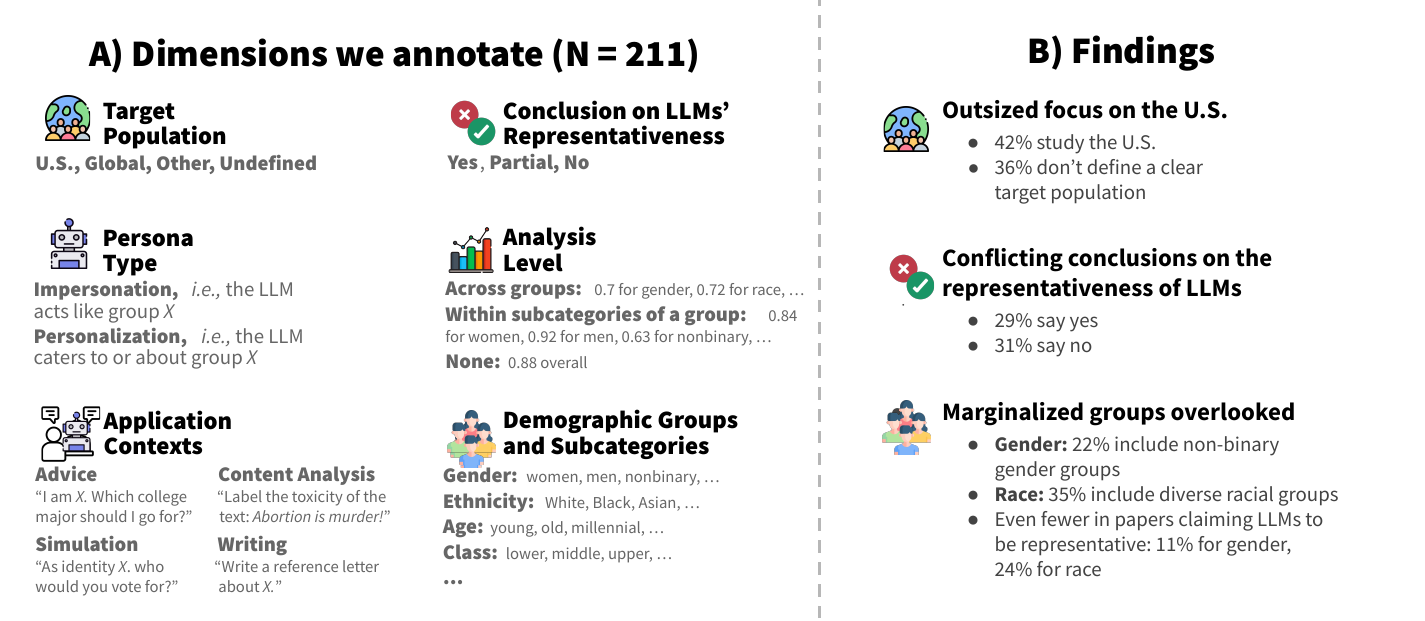}
    \caption{Description of the (A) dimensions we annotate for each paper in this review  and (B) key findings related to the demographic representativeness of LLMs.}
    \label{fig:fig1}
\end{figure*}


In the emerging field of social applications of LLMs, current studies reach opposing conclusions on their demographic representativeness, even when analyzing the same populations using similar techniques and models. For example, while~\citeauthor{argyle2023out} find that LLMs represent American populations via prompt-induced personas, ~\citeauthor{santurkar2023whose} conclude that LLMs only reflect the opinions of certain groups. Additionally, other researchers find that LLMs reduce the variance of behavior within groups and flatten~\cite{wang2024largelanguagemodelsreplace} and caricature people~\cite{cheng2023compost}. While literature surveys on algorithmic bias in LLMs exist~\cite{gupta2024sociodemographic,gallegos2024biasfairnesslargelanguage,chu2024fairnesslargelanguagemodels}, as well as a scoping review on LLMs supplementing humans in human-subject studies~\cite{agnew2024illusion}, none comprehensively review the fine-grained demographic dimensions probed in social applications of LLMs and their connection to representativeness. 

Motivated by this lack of systematic literature analyses, we survey 211 articles across a variety of LLM applications, asking and answering the following research questions: \textbf{RQ1: Which demographic dimensions are probed, and in which contexts?}\footnote{We use the term `demographic' to include both demographic and sociodemographic groups.} We investigate the demographic categories studied and evaluated in these papers, as well as how these categories are operationalized, including which subcategories are considered (Figure~\ref{fig:fig1}A). We then assess \textbf{if there is consensus on the representativeness of LLMs? (RQ2)}.



\textbf{Findings:} While the majority of papers find no evidence for representativeness or that LLMs are \textit{partially} representative for a certain group only, 29\% claim that LLMs are representative. Assessing potential causes of this divergences, we find that among papers claiming LLMs to be representative, 30\% do not evaluate representativeness across multiple demographic groups, nor across the subgroups of a particular demographic. Instead these papers report overall LLM representativeness. Another 35\% of these papers make conclusions about gender representativeness without reporting the gender subcategories they study. The equivalent proportion for racial categories is even higher. This type of underreporting on demographic factors is comparatively lower in studies that claim partial or no representativeness. While studies with a negative outlook also include a higher portion of marginalized racial and gender categories in their study, only around a fifth of all studies (22\%) include subcategories beyond the gender binary.
Finally, most studies either explicitly or implicitly focus on people in the U.S. excluding other relevant (sub)populations. 

\textbf{Contributions. }Our work contributes a systematic understanding of the state-of-the-art research on the demographic representativeness of LLMs, finding patterns of underreporting of crucial details required to establish representativeness. We provide a set of specific recommendations for future research on this topic for better documentation and evaluation.

\section{Background}

\textbf{Representativeness in NLP and Beyond.} \citeauthor{chasalow2021} describe representativeness, like bias, as a `suitcase word'—a term used widely but with multiple definitions. In quantitative social sciences, representative samples allow studying large populations without surveying every member~\cite{gobo2004sampling}. In Computational Linguistics, NLP, and LLMs, it refers to ``the extent to which a sample includes the full range of variability in a population''~\cite{biber1993representativeness}. In sociolinguistics, representativeness is often linked to generalizing across languages and varieties~\cite{grieve2025sociolinguistic}.

There are extensive studies of algorithmic bias in NLP, including sociodemographic bias~\cite{gupta2024sociodemographic}. Bias has many definitions, some of which often focus on misrepresentation or underrepresentation of certain (demographic) groups~\cite{ferrara2023fairness}. Recent research has also focused on `AI Alignment', which is broadly defined as making ``AI systems behave in line with human intentions and value''~\cite{ji2023ai}. There are several parallels between alignment and representativeness, especially for personalized LLM agents; however such parallels have not been widely explored, possibly due to the lack of a concrete vocabulary for operationalizing alignment~\cite{kirk2023empty}. 

\textbf{Repurposing Bias for Representativeness? }\citeauthor{argyle2023out} conceptualize `algorithmic fidelity', positing that biases in LLMs conditioned on demographic attributes can mirror ideas and opinions of those demographics. They state:
\textcolor{gray}{\textit{``... the ``algorithmic bias'' within one such tool—the GPT-3 language model—is instead both fine-grained and demographically correlated, meaning that proper conditioning will cause it to accurately emulate response distributions from a wide variety of human subgroups.''}} However, algorithmic bias in NLP systems, including LLMs, often has competing definitions, with some emphasizing underrepresentation and misrepresentation as key factors~\cite{ferrara2023should,gallegos2024biasfairnesslargelanguage}. Given its multifaceted nature, can bias enhance equitable representativeness across all groups, including marginalized ones?\footnote{Studies affirming LLMs' algorithmic fidelity, e.g.,~\citeauthor{argyle2023out,kim2023ai}, do not explicitly define bias.} In this review, we attempt to find the current consensus w.r.t to the demographic representativeness of LLMs and unearth potential reasons behind seemingly conflicting findings.

\begin{table}[]
\small
\centering
\begin{tabular}{@{}llll@{}}
\toprule
\textbf{Source}                                                        & \textbf{Longlisted} & \textbf{Deduplicated} & \textbf{Included} \\ \midrule
\citeauthor{agnew2024illusion}                                            & 13         & 13           & 4        \\
ArXiv                                                         & 291        & 290          & 156      \\
\begin{tabular}[c]{@{}l@{}}ACL\end{tabular}       & 196        & 41           & 9       \\
\begin{tabular}[c]{@{}l@{}}Semantic\end{tabular}    & 86         & 4            & 1        \\
\begin{tabular}[c]{@{}l@{}}ACM DL\end{tabular} & 117        & 108          & 5        \\
OpenAlex                                                      & 362        & 160          & 29       \\
\begin{tabular}[c]{@{}l@{}}Other\end{tabular} & 24         & 12            & 7        \\
\midrule
Total                                                         & 1076       & 615          & 211      \\ \bottomrule
\end{tabular}
\caption{Summary statistics of papers considered in this review.}
\label{tab:stats}
\end{table}

\begin{table*}[t!]
\centering
\footnotesize
\begin{tabular}{@{}lll@{}}
\toprule
\textbf{Category}                                                                                        & \textbf{Subcategory}                                                & \textbf{Definition and Examples}                                                                                                                                                                                                                                                                                                                                                      \\ \midrule
\multirow{1}{*}{Contexts}                                                                       & Advice                                                     & \begin{tabular}[c]{@{}l@{}}Providing help with decision-making, or giving suggestions, recommendations,\\ or advice, e.g.,~\cite{levy2024evaluating,10.1145/3689904.3694709,lahoti2023improving}\end{tabular}                                                                                                                                                                                                                                    \\ \cmidrule(l){2-3}
                                                                                                & Simulation                                                 & \begin{tabular}[c]{@{}l@{}}Synthetic data generation to study human behavior directly, e.g., simulating survey\\ respondents \cite{bisbee2024synthetic} or platform simulations~\cite{park2022social}.\end{tabular}                                                                                                                                                               \\ \cmidrule(l){2-3}
                                                                                                & \begin{tabular}[c]{@{}l@{}}Content\\Analysis\end{tabular} & \begin{tabular}[c]{@{}l@{}}Qualitative content labeling, evaluation, and labeling, e.g., sentiment analysis, \\ hate speech~\cite{beck2024sensitivity,giorgi2024human,sun2023aligningwhomlargelanguage}\end{tabular}                                                                                                                                                                                                  \\ \cmidrule(l){2-3}
                                                                                                & Writing                                                    & \begin{tabular}[c]{@{}l@{}}Fiction or non-fiction writing, could also include translation or rewriting content\\ e.g., ~\cite{wan2024whitemenleadblack,sourati2024secretkeepersimpactllms}\end{tabular}                                                                                                                                                                                                                                       \\ \cmidrule(l){2-3}
                                                                                                & \begin{tabular}[c]{@{}l@{}}Generic\end{tabular}    & \begin{tabular}[c]{@{}l@{}}General investigation of LLMs, without any downstream context e.g.,\\~\cite{zhao2023gptbias,jiang2022communitylm}\end{tabular}.                                                                                                                                                                                                                                                                                                            \\
                                                                                                \midrule
\multirow{2}{*}{\begin{tabular}[c]{@{}l@{}}Persona\\ Type\end{tabular}}                  & Impersonation                                                      & \begin{tabular}[c]{@{}l@{}}Persona induced in the LLM, e.g., “answer this question as a \textit{Democrat}” using \\personas from survey data in~\citet{argyle2023out,vonderheyde2024voxpopulivoxai}\end{tabular}                                                                                                                                                                                                        \\ \cmidrule(l){2-3}
                                                                                                & Personalization                                                     & \begin{tabular}[c]{@{}l@{}}Persona that the LLM needs to act upon, e.g., text written by a group ``would you\\hire this \textit{man} based on his resume?''~\cite{gaebler2024auditing} or about a group, e.g.,\\ targets of hate speech ``annotate: [content targeting \textit{women}]''~\cite{beck2024sensitivity}\end{tabular}                                                                                                                                                                            \\ \midrule
\multirow{3}{*}{\begin{tabular}[c]{@{}l@{}}Conclusion on\\ Representative\\ -ness\end{tabular}} & Yes                                                        & \begin{tabular}[c]{@{}l@{}}The study is positive, e.g.~\citeauthor{argyle2023out}, who say \textit{\textcolor{gray}{"We suggest that language}}\\\textit{\textcolor{gray}{models with sufficient algorithmic fidelity thus constitute a novel and powerful}}\\\textit{\textcolor{gray}{tool to advance understanding of humans and society across a variety of disciplines."}}\end{tabular}                                                                             \\ \cmidrule(l){2-3}
                                                                                                & Partial                                                    & \begin{tabular}[c]{@{}l@{}}The study has mixed results, finding LLMs to be successful at representating\\ some groups but not others, e.g.~\citeauthor{gabriel2024can}, who say \textit{\textcolor{gray}{``We find that while GPT-4}}\\ \textit{\textcolor{gray}{can reflect and amplify harmful biases found in peer-to-peer support, these biases}}\\ \textit{\textcolor{gray}{vary significantly based on prompt design and can be mitigated through...}}"\end{tabular} \\ \cmidrule(l){2-3}
                                                                                                & No                                                         & \begin{tabular}[c]{@{}l@{}}The study has a negative outlook on representativeness, e.g.,~\citeauthor{vonderheyde2024voxpopulivoxai}\\ noting \textit{\textcolor{gray}{"We have shown that in its current state, GPT-3.5 is not suitable for}}\\ \textit{\textcolor{gray}{estimating public opinion across (sub)populations..."}}\end{tabular}                                                                                                            \\ 
                                                                                                \bottomrule
\end{tabular}
\caption{\textbf{Key categories, definitions, and examples in our annotation codebook.} An expanded version of the codebook with all categories and the instructions given to annotators are included in the Appendix (Section \ref{sec:app_codebook}).}
\label{tab:codebook}
\end{table*}
\section{Literature Search and Annotation}

In this literature review, we are interested in the intersection of LLMs and demographics; as such we only include papers that conduct a study which incorporates demographics somewhere in the pipeline, i.e., either use demographic dimensions in input to LLMs or include demographic variables in the evaluation of LLMs. Therefore, we search for papers containing the keywords ``Large Language Models''/``LLM'' and ``demographic*'' available online before December 1st, 2024. We first started with the 13 papers assessed in the scoping review by~\citeauthor{agnew2024illusion} on the potential of replacing human participants with LLMs in human-subject studies.\footnote{Out of the 16 artifacts studied in~\citeauthor{agnew2024illusion}, there are 13 research papers, while the other 3 are product offerings.} To expand this list, we utilize five sources --- \citeauthor{arxiv}, the \citeauthor{aclanthology}, \citeauthor{semantic_scholar}, \citeauthor{acm_dl}, and \citeauthor{openalex}. The latter is an open-source version of the Microsoft Academic Graph. Finally, we also include existing community resources, i.e., papers identified in~\citeauthor{simmons2023largelanguagemodelssubpopulation} and a paper list on public opinion simulation with LLMs.\footnote{\url{https://github.com/CaroHaensch/public_opinion_llms}} 
After a semi-automatic deduplication step, three authors split the 615 papers between them to manually assess whether a paper should be included in the literature review. 

\subsection{Scope and Inclusion Criteria}
We restrict our literature review to research papers (not necessarily peer-reviewed) with empirical findings. As such, we exclude other literature reviews, perspective and theoretical articles, pay-walled articles, and extended abstracts (but include short papers and workshop papers). The first content-related criterion for inclusion is that the study should touch on demographics. Only four of the 13 papers studied in~\citeauthor{agnew2024illusion} do so~\cite{argyle2023out,park2022social,aher2023using,gerosa2024can}; the other nine discuss the potential of LLMs replacing humans, but do not state \textit{which} humans. 
To balance coverage of relevant literature with the annotation workload, we only include generative LLM-based studies which are text-only, i.e., excluding vision, speech, or multimodal applications. Based on these criterion, we include 211 papers. The literature search and inclusion process is summarized in Table~\ref{tab:stats}.

\subsection{Codebook Categories}

We have a three-part annotation scheme whose most important categories of the codebook are exemplified in Table~\ref{tab:codebook}, while the full codebook can be found in the Appendix (Section~\ref{sec:app_codebook}).

\textbf{Contexts and LLMs. }Contexts refer to the scenario in which LLMs are used, either as proxies for humans or for providing services to or about humans. These categories are based on how people use LLMs~\citep{mireshghallahtrust}, restricted to those where demographics play an explicit role. In addition, in line with~\citeauthor{tseng-etal-2024-two}, we define and annotate two types of representativeness or \textit{personas} in LLMs: their ability to \textit{impersonate} group \textit{X} and their ability to \textit{personalize} to group \textit{X}. We also note the LLMs used and the approach to induce or improve their representativeness, e.g., prompting.

\textbf{Evaluating and Improving Representativeness.} We annotate the \textbf{response format} employed in each study, i.e., free-text responses from LLMs vs. closed-form responses, like multiple choice question-answering (QA). 
Except~\citeauthor{lee2025enhancing} who study LLMs' ability to adapt to African American English, all papers in our sample focus on broad and/or multiple demographic categories. To establish an LLM's equitable representativeness for a population, it is important to understand how well´it represents different subgroups of that population. Therefore, we label whether studies conduct a \textbf{demographically disaggregated evaluation}, i.e., if they evaluate a model \textit{across} multiple groups, \textit{within} subgroups of demographic category, or both. 

\begin{figure*}[t!]
    \centering
    \subfloat[\centering Context vs. Persona Type]{{\includegraphics[width=.95\linewidth]{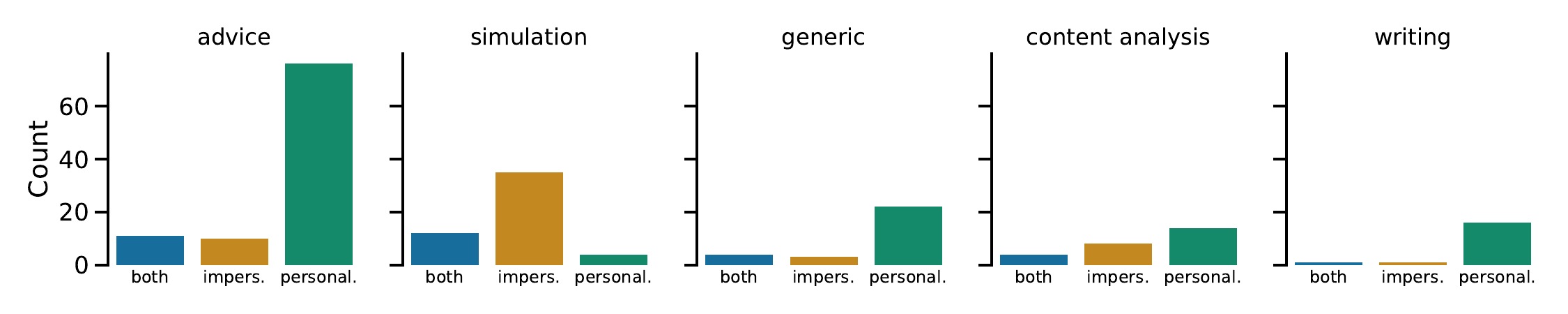} }}
    \qquad
    \subfloat[\centering Context vs. Response Format]{{\includegraphics[width=.95\linewidth]{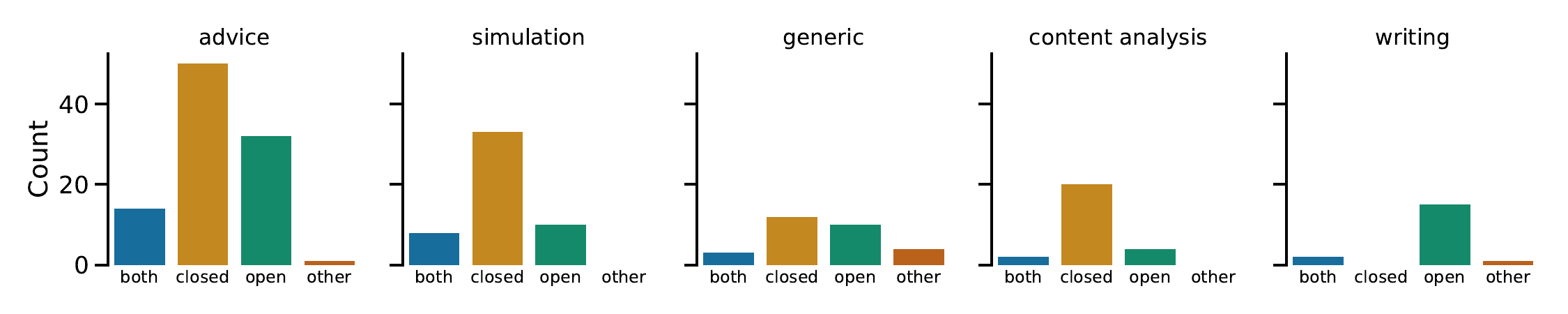} }}
    \caption{\textbf{Distribution of contexts over a) persona type and b) response format.} Most papers look into \textit{personalization} of LLMs across different contexts, except \textit{simulation} studies. Closed-form responses from LLMs are more widely studied except in \textit{writing}. 
    }%
    \label{fig:context_descriptive}
\end{figure*}

\begin{table}[]
\footnotesize
\begin{tabular}{@{}lll@{}}
\toprule
\textbf{Paper}                              & \begin{tabular}[c]{@{}l@{}}\textbf{Dem.}\\ \textbf{Category}\end{tabular} & \begin{tabular}[c]{@{}l@{}}\textbf{Subcategories} and\\ \textbf{Descriptors}\end{tabular}                                                                                                                                \\ \midrule
\multirow{3}{*}{\citeauthor{zheng2024dissecting}}      & Gender                                                         & \begin{tabular}[c]{@{}l@{}}Male, Female, \\ LGBTQ\end{tabular}                                                                                                                                        \\ \cmidrule(l){2-3} 
                                   & Race                                                           & \begin{tabular}[c]{@{}l@{}}American Indian or\\ Alaskan Native,\\ Asian, Black or \\ African American, \\ Filipino, Hispanic\\ or Latino, Native\\ Hawaiian or Pacific\\ Islander, White\end{tabular} \\ \cmidrule(l){2-3} 
                                   & Income                                                         & \begin{tabular}[c]{@{}l@{}}Disadvantaged, \\ Non-disadvantaged\end{tabular}                                                                                                                           \\ \midrule
\multirow{2}{*}{\citeauthor{alipour2024robustness}}     & Gender                                                         & Man, Woman                                                                                                                                                                                            \\ \cmidrule(l){2-3} 
                                   & Race                                                           & \begin{tabular}[c]{@{}l@{}}Asian, Black, White,\\ Hispanic\end{tabular}                                                                                                                               \\ \midrule
\multirow{3}{*}{\citeauthor{steinmacher2024can}} & Gender                                                         & \multirow{3}{*}{Not reported}                                                                                                                                                                         \\ \cmidrule(lr){2-2}
                                   & Location                                                       &                                                                                                                                                                                                       \\ \cmidrule(lr){2-2}
                                   & Age                                                            &                                                                                                                                                                                                       \\ \bottomrule
\end{tabular}
\caption{Examples of how demographic categories and subcategories are operationalized and reported in papers.}
\label{tab:demographic_examples}
\end{table}

\textbf{Demographics and Representativeness. }We identify the demographic categories studied in a paper and how they are operationalized, i.e. the \textbf{subcategories} and \textbf{descriptors} used to represent these subgroups (examples in Table~\ref{tab:demographic_examples}).\footnote{The full list of demographic categories we include in this paper can be found in Table~\ref{tab:app_demographic_categories} in the Appendix.} Subcategories refer to subgroups of a given demographic category, while descriptors refer to how these subgroups are described, e.g,~\citeauthor{argyle2023out} use the binary gender descriptors `male' and `female' vs. ~\citeauthor{cheng2023marked} who use `man', `woman', and `nonbinary', while~\citeauthor{deldjoo2024understanding} does not specify the gender subcategories or descriptors used. 
We also annotated the \textbf{target population} studied in a paper, e.g., the global population in~\citeauthor{durmus2023towards}. While some papers do not explicitly specify a target population, for some of them we can infer whether the population is the U.S. based on racial and political leaning descriptors, i.e., using U.S. specific-categories like `Native American' or `Republican', e.g.,~\citeauthor{arzaghi2024understanding}.

Finally, we annotate the paper's \textbf{conclusion regarding the representativeness of LLMs} for the target population studied. 
Many of the papers included in our survey focus on mitigating demographic biases in LLMs through debiasing and alignment, e.g.~\citet{do-etal-2025-aligning,he2025cosenhancingpersonalizationmitigating}. 
We annotate these articles as concluding positively if they find their debiasing technique to be effective, as biases are considered a threat to representativeness~\cite{ferrara2023fairness} and reducing them would lead to improved representativeness. 

\subsection{Annotation Process}

Three annotators, who are also authors of this paper, independently annotated all 211 papers using the aforementioned codebook over three rounds. After each round, checks were done by randomly selecting three papers from each annotator's batch to be annotated by all three annotators to ensure reliability. We found little disagreement across three rounds (3-8\% of diverging annotations across the rounds). 
Disagreements were resolved after discussion (c.f. Appendix Section~\ref{app_annotation} for more details.) 

\begin{figure*}[t!]
    \centering
    \includegraphics[width=0.99\textwidth]{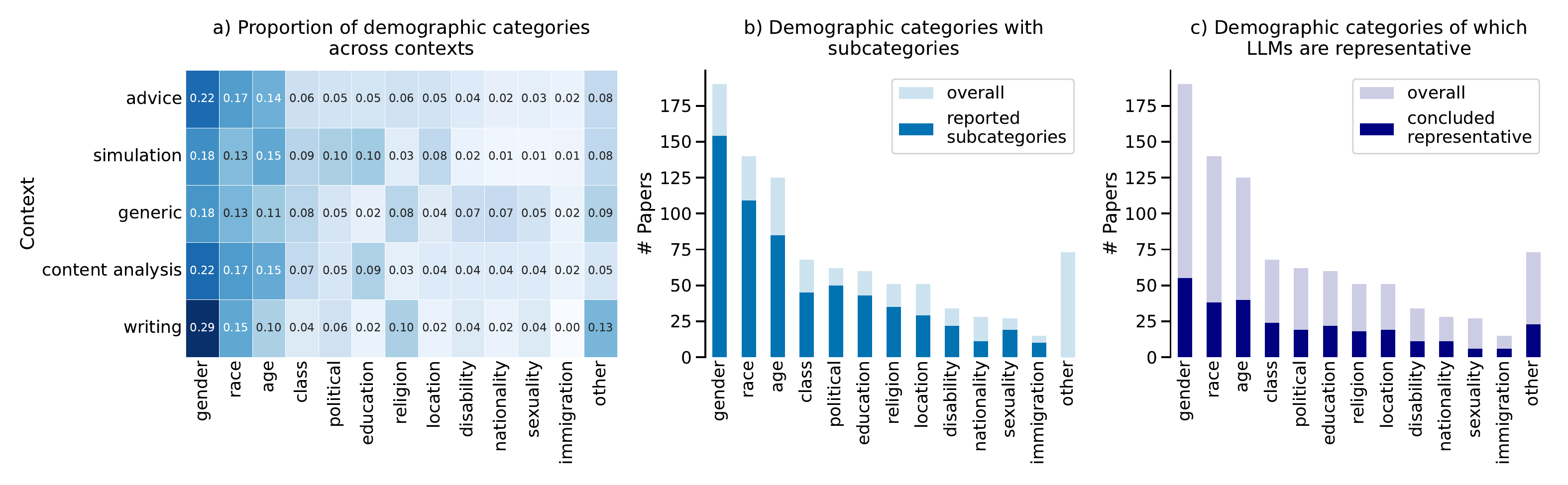}
    \caption{\textbf{Demographic Dimensions a) proportionally across contexts, b) with subcategories, and c) concluded to be represented by LLMs.} \textit{Gender} and \textit{Race} are widely studied, but \textit{simulation} tends to have comparatively more balanced distribution of demographics. }
    \label{fig:demo_stats}
\end{figure*}

\section{Results}

\subsection{Descriptive Findings}
A majority of studies fall under \textit{advice} (43\%), followed by \textit{simulation} (23\%), \textit{generic} (13\%) and \textit{content analysis} (11\%). \textit{Advice} scenarios span many different topics including medical~\cite{rawat2024diversitymedqa}, hiring~\cite{gaebler2024auditing}, and education~\cite{weissburg2024llms}. Simulations often focus on replicating surveys~\cite{gerosa2024can} or social media behavior~\cite{chang2024llms}. \textit{Content analysis} studies often focus on annotating subjective tasks with LLMs~\cite{jiang-etal-2024-examining}.\footnote{Some have multiple contexts (c.f. Appendix~\ref{app_annotation}), while~\citeauthor{mori2024algorithmicfidelitymentalhealth} use LLMs for the sole purpose of generating training data. As this was the only paper for this application, it did not justify creating a new context. We therefore exclude it from the analysis on contexts.} Figure~\ref{fig:context_descriptive}a shows that across most contexts, \textit{personalization} is more common, except \textit{simulation} where \textit{impersonation} or both are studied. Figure~\ref{fig:context_descriptive}b shows the distribution of response formats across contexts; besides \textit{content analysis} where mainly closed evaluation is conducted and the opposite in \textit{writing}, we see both open and closed in other contexts, with closed format being more prevalent. 

\textbf{LLMs and Methods for Steering them. }While, most studies (62\%) include more than one model, a majority of studies only use commercial LLMs; 80\% of the papers include at least one OpenAI model. This is followed by open-weight models like LLaMa (39\%), and Mistral or Mixtral (21\%).\footnote{To keep the annotation and analysis from blowing up, we do not report parameter size or versions of LLMs used. More descriptive results can be found in Appendix~\ref{app:more_descriptive}.} 
Finally, in terms of measuring and steering representativeness, prompt-based techniques are by far the most commonly used, appearing in 64\% of the papers, followed by fine-tuning approaches (13\%), e.g.,~\citet{jiang2022communitylm,wald2023exposing}.

\subsection{RQ1: Which demographics are most studied and in which contexts?}
We first investigate the target populations mentioned in papers (Table~\ref{tab:target_pop}). A large number of papers, i.e. 36\%, do not specify a target population, instead aiming to study demographic characteristics in general with generic categories, e.g., white vs. non-white in~\citeauthor{kamruzzaman2024woman} or do not mention either an explicit target population or demographic subcategories~\cite{do-etal-2025-aligning}. 26\% of the papers solely and explicitly focus on the U.S., while another 16\% do so implicitly via the use of U.S.-specific racial or political subcategories.
18\% of studies explicitly mention target populations beyond the U.S., while a small proportion (4\%) attempt to study the global population. 

Figure~\ref{fig:demo_stats}a shows the proportion of demographic dimensions studied over different contexts, while Figure~\ref{fig:demo_stats}b and c show the count of different demographic dimensions. We confirm previous findings in NLP research showing gender and racial categories to be the most studied~\cite{gupta2024sociodemographic}, though the distribution is less skewed compared to research on biases~\cite{gupta2024sociodemographic}. We also note that for \textit{simulation} studies, the distribution is more balanced compared to other contexts. Widely studied categories in `other' include marital status, number of children, and occupation.

\begin{table}[]
\footnotesize
\begin{tabular}{@{}llll@{}}
\toprule
\begin{tabular}[c]{@{}l@{}}\textbf{Target}\\\textbf{Pop.}\end{tabular} & \begin{tabular}[c]{@{}l@{}}\textbf{\#}\\\end{tabular} & \textbf{Examples}                                                                                                                                     & \begin{tabular}[c]{@{}l@{}}\textbf{\% Repres}\\\textbf{-entative?}\end{tabular} \\ \midrule
\begin{tabular}[c]{@{}l@{}}Other\end{tabular}                                                        & 39        & \begin{tabular}[c]{@{}l@{}}German Political\\Parties (\citeauthor{batzner2024germanpartiesqa}),\\Indians (\citeauthor{sahoo2024indibias})\end{tabular}   & 23\%                                                                              \\
\midrule
\begin{tabular}[c]{@{}l@{}}U.S. Explicit\end{tabular}                                                & 54        & \begin{tabular}[c]{@{}l@{}}\citeauthor{argyle2023out},\\\citeauthor{santurkar2023whose}\end{tabular}   & 22\%                                                                              \\
\midrule
\begin{tabular}[c]{@{}l@{}}U.S. Implicit\end{tabular}                                                & 34        & \begin{tabular}[c]{@{}l@{}}\citeauthor{cheng2023compost},\\\citeauthor{giorgi2024human}\end{tabular}   & 14\%                                                                              \\
\midrule
\begin{tabular}[c]{@{}l@{}}Global\end{tabular} & 9         & \begin{tabular}[c]{@{}l@{}}\citeauthor{durmus2023towards},\\ \citeauthor{jin2024language}\end{tabular}   & \textcolor{red}{11\%}                                                                               \\
\midrule
Undefined                                                    & 75        & \begin{tabular}[c]{@{}l@{}}\citeauthor{park2023generative},\\\citeauthor{lahoti2023improving}\end{tabular} & \textcolor{blue}{45\%}                                                                              \\
\bottomrule
\end{tabular} 
\caption{\textbf{Target Populations and their proportion found to be represented by LLMs.} Studies on Global populations report the lowest rate (11\%). 
}
\label{tab:target_pop}
\end{table}

In terms of how specific demographic categories are operationalized, we find that many papers (38\% on average across all demographics) do not explicitly report the demographic subcategories and descriptors they use in their study, i.e., they state that they study a particular demographic dimension but do not report the full list of subcategories and/or descriptors (Figure~\ref{fig:demo_stats}b).\footnote{Note that reporting descriptors does not apply to papers that do not prompt LLMs with demographic personae, but reporting subcategories does.} We note that studies focusing on \textit{nationality} and \textit{class} tend to underreport the subcategories used more compared to other categories. The findings from our analysis of the target population and demographics studied reveal two patterns across most LLM application contexts --- \textbf{i) an outsized focus on the U.S. population}, in line with previous research~\cite{field2021survey} and persistent in studies with LLMs, and ii) \textbf{a tendency to under-specify the explicit target population and demographic subgroups being studied.}

\subsection{RQ2: Is there consensus on the representativeness of LLMs?}\label{sec:consensus}

\begin{figure}
    \centering
    \includegraphics[width=0.99\linewidth]{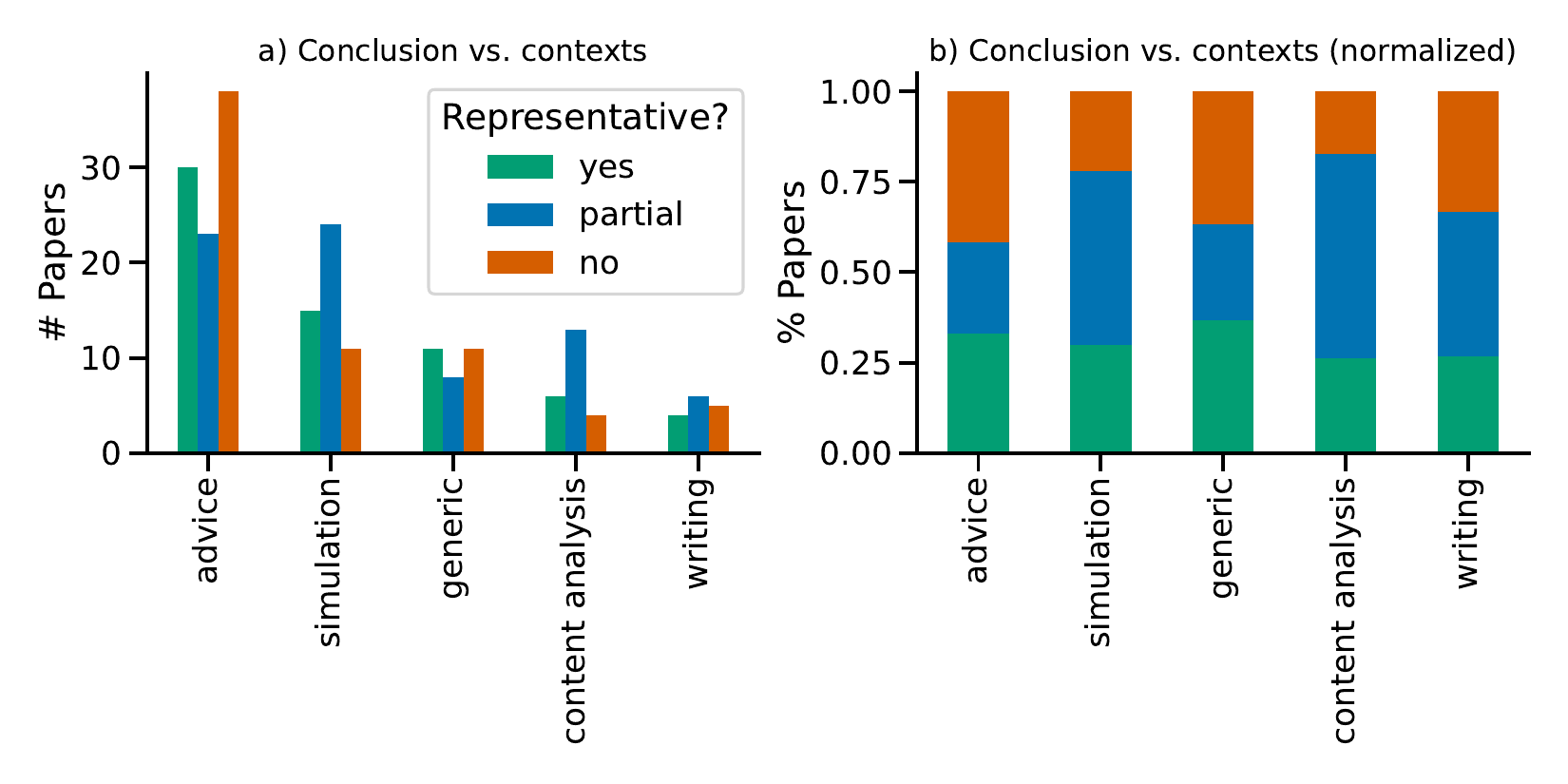}
    \caption{\textbf{Contexts vs. Representativeness.} While \textit{advice} and \textit{generic} have polarized responses, the rest report mainly partial representativeness.}
    \label{fig:context_conclusion}
\end{figure}

Out of 211 papers, 29\% conclude `yes' on representativeness of LLMs, 34\% conclude `partial', and 32\% conclude `no'.\footnote{11 papers or 5\% do not provide a conclusion, hence we exclude them from the analysis in Sections~\ref{sec:consensus} and \ref{sec:deepdive}.} Figure~\ref{fig:context_conclusion}a and Figure~\ref{fig:context_conclusion}b shows the distribution and proportion of studies claiming representativeness of LLMs across different contexts, respectively. 
\textit{Advice} and \textit{generic} studies seem to have strongly diverging conclusions on representativeness while \textit{simulations}, \textit{writing}, and \textit{content analysis} have a majority of partially representative findings. For the latter context's majority partial results, previous findings on the limits of demographic dimensions in subjective data annotation~\cite{orlikowski2023ecological}, appears to extend to LLMs' annotations~\cite{beck2024sensitivity,alipour2024robustness}. With respect to steering, 24\% of the papers using prompting to induce personas conclude positively, while the number is much higher for fine-tuning (63\%).

In terms of demographic factors, studies that implicitly target the US population or a global population have the lowest percentage of conclusions on the positive representativeness of LLM (Table~\ref{tab:target_pop}, last column). However notably, studies which do not specify a target population had the most positive conclusions. Many of them are on debiasing~\cite{lahoti2023improving} or alignment~\cite{chen2024spicaretrievingscenariospluralistic}. However, it is unclear to which exact populations these findings apply. In Figure~\ref{fig:demo_stats}c, we study the number of papers found to be representative across different demographic categories. This distribution generally mirrors the rank of demographic dimensions studied, however \textit{political leaning}, \textit{disability}, and \textit{sexuality} have comparatively fewer studies claiming that LLMs are representative. 

\begin{figure}
    \centering
    \includegraphics[width=0.95\linewidth]{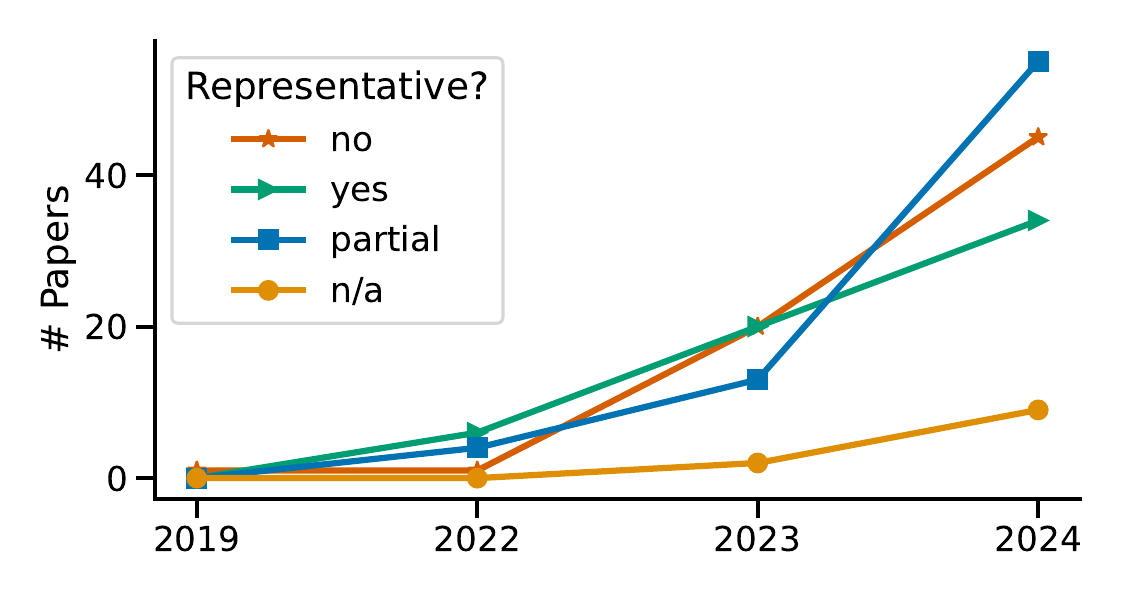}
    \caption{\textbf{Reported representativeness of LLMs over time.} We can observe (i) an increase of conflicting reports and (ii) an especially high increase in papers finding partial representativeness.}
    \label{fig:temporal}
\end{figure}

Finally, in Figure~\ref{fig:temporal}, we investigate trends related to LLM representativeness over time and note two points --- a relatively slower growth of articles with a positive outlook and an increase in papers claiming partial representativeness, suggesting a move to more nuanced evaluations.

\begin{figure*}
    \centering
    \includegraphics[width=0.99\linewidth]{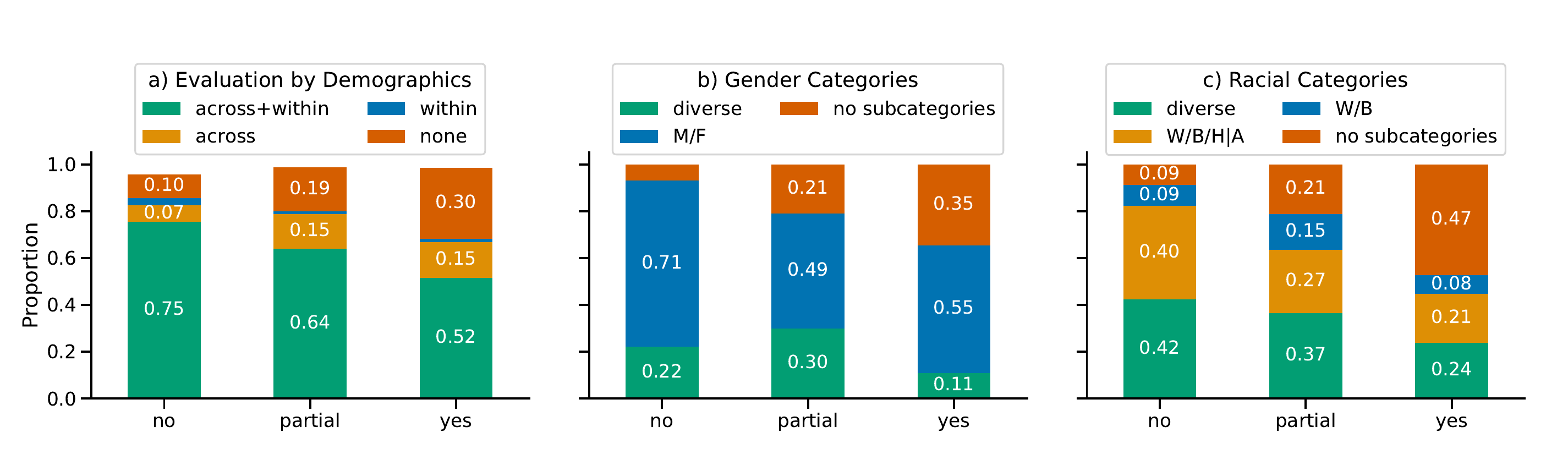}
    \caption{\textbf{Factors differentiating studies claiming LLMs are representative vs. those claiming otherwise.} We plot the proportion of papers conducting a) demographically disaggregated analysis, and the proportion of papers studying different types of b) gender and c) racial categories. We find that papers with a positive outlook show a higher tendency of not conducting any demographically disaggregated analysis. They also more often underreport demographic subcategories or do not include marginalized/diverse categories.}
    \label{fig:disagreements}
\end{figure*}

\subsection{Disentangling Disagreements on Representativeness}\label{sec:deepdive}

To unpack disagreements regarding outlooks on representativeness of LLMs, we assess the relationship between these outlooks across a) evaluation approaches and b) demographic categories studied.

\textbf{Demographically Disaggregated Results. }\textcolor{black}{Overall, we find that 20\% of the papers do not conduct any type of demographically disaggregated analysis, i.e., they report the LLMs' overall performance rather than performance across multiple demographic groups or within the subcategories of a group.} This proportion is higher for papers claiming LLMs to be representative, i.e., 30\% compared to 19\% of the papers claiming LLMs to be partially representative and 10\% claiming no representativeness ( Figure~\ref{fig:disagreements}a).\footnote{Papers which only focus on a single demographic are included under `across + within' in Figure~\ref{fig:disagreements}a if they report results within the subcategories of that demographic dimension.} 

\textbf{Demographic Categories. }Building on findings in past research on the tendency of LLMs to stereotype marginalized groups~\cite{cheng2023compost,nguyen2024simulating}, we assess whether including marginalized or diverse categories is associated with findings on representativeness. 

We investigate the two most studied demographic categories, \textit{gender} and \textit{race}, and investigate which, if any, descriptors have been used to operationalize these two demographic dimensions. For both \textit{gender} and \textit{race}, we devise a category for no reported subcategories and, as a consequence, no descriptors. For \textit{gender}, we have one category encoding binary gender (`male/female') and another category including diverse categories (`diverse') if a study includes any gender subcategory besides `male' or `female'. Similarly for \textit{race}, we define a `Black/White' category, `White/Black/Hispanic or Asian' if either Asian or Hispanic is included with White and Black,\footnote{We do not find any studies that use `Asian' or `Hispanic' without also including `White' and `Black'.} and finally a `diverse' racial category if the target population is explicitly beyond the U.S. (i.e., Global or Other) or if they include the identities `Native American' or `Pacific Islander'. 

Figure~\ref{fig:disagreements}b and~\ref{fig:disagreements}c show the proportion of the aforementioned categories across studies claiming representativeness of LLMs. \textbf{For both \textit{gender} and \textit{race}, many studies claiming LLMs to be representative either do not report the demographic subcategories or have a lower proportion of diverse categories compared to studies claiming otherwise.} For example, from studies that claim that LLMs are representative of racial demographics, 47\% do not report racial subcategories, compared to 9\% in studies claiming no representativeness. Our findings across several papers strengthens the hypothesis that LLMs might be particularly unrepresentative of marginalized groups, e.g.,~\citeauthor{argyle2023out,kim2023ai} use binary gender and find LLMs to be capable of simulating the U.S. population, while~\citeauthor{cheng2023compost,wang2024largelanguagemodelsreplace} investigate nonbinary personas as well, coming to the opposite conclusion. Similarly, for \textit{advice}, only three of 30 papers concluding positively about the representativeness of LLMs, use diverse \textit{gender} subcategories; they aim to debias~\cite{lahoti2023improving,tamkin2023evaluating} or align existing non-representative LLMs~\cite{li2024steerability}. 
However, we also note that while papers claiming LLMs to be representative tend to exclude marginalized groups more, the inclusion of these groups is generally low overall --- 22\% and 35\% of all papers include diverse gender and racial subgroups, respectively. This indicates a greater need for studies on the representativeness of LLMs for marginalized groups.

\textbf{Other Factors. }Conducting a qualitative analysis of other factors driving disagreement, we find that many papers concluding positively rely more on closed-form response formats or often do not take into account the variance of LLM responses. More details are provided in Appendix~\ref{sec:other}.


\section{Discussion}

LLMs, beyond their role as assistive chatbots, are increasingly used to supplement or replace humans in research \cite{gilardi2023chatgpt}. In all these applications, it is crucial to assess whether LLMs provide equitable assistance and adequately represent the populations they aim to supplement. Some studies (\citet{argyle2023out,kim2023ai} \textit{inter alia}) argue that LLM biases enhance subgroup representation, while others highlight contradictions between bias and representativeness~\cite{ferrara2023should,wang2024largelanguagemodelsreplace}. Empirical studies remain divided on LLM representativeness \cite{argyle2023out,bisbee2024synthetic}. Our systematic review finds that studies incorporating demographics predominantly focus on the U.S. and often lack crucial details to assess representativeness.

\subsection{Implications for using LLMs in Social Applications}
LLMs' flexibility grants researchers broad methodological choices—e.g., persona induction (prompting vs. fine-tuning), response types, and model selection. These issues contribute to growing concern regarding reproducibility, even without factoring in the reproducibility issues associated with using closed-source models~\cite{barrie2024replication}. Our review shows that these degrees of freedom do not just affect the assessment of reproducibility, but also of representativeness. For instance, \citeauthor{argyle2023out} and \citeauthor{bisbee2024synthetic} reach opposing results despite similar methodologies. Furthermore, marginalized groups remain underrepresented and underserved by LLMs~\cite{wang2024largelanguagemodelsreplace,cheng2023compost}. Therefore, studies assessing LLMs for social applications should report their exact design choices, \textbf{while evaluating the representativeness of LLMs across multiple subgroups within the target population, rather than relying on an overall assessment.} 
\subsection{Recommendations for Reporting and Improving Representativeness}

While many papers anticipate future LLM improvements in representativeness, the exact changes needed remain unclear. To gauge these improvements, context-specific evaluations are essential. Current LLM benchmarks like HELM~\cite{liang2022holistic} or BigBench~\cite{srivastava2022beyond} include some bias-related evaluations, however, in line with our findings, bias and representativeness are not necessarily equivalent. We advocate for tailored benchmarks explicitly defining target populations and demographic subcategories, along with demographically disaggregated analyses combining open and closed-form evaluations. To enhance transparency, we propose incorporating explicit population and demographic categories into reproducibility checklists and model/data documentation. 

Additionally, future research should explore under-examined representativeness interventions, including model editing and RLHF. Algorithmically biased LLMs might only represent a particular group of people --- non-marginalized Americans in line with previous findings~\cite{durmus2023towards,atari2023humans}, in narrow evaluation settings. To represent diverse populations, we need to move beyond repurposing algorithmic bias and think of intentionally designed representative LLMs, e.g., through approaches like more detailed personas~\cite{moon-etal-2024-virtual} or pluralistic alignment~\cite{sorensen2024position}. In data-driven approaches, robust sampling strategies from quantitative social sciences can also inform better demographic representation, especially for marginalized populations~\cite{freimuth1990there}. However, technical solutions might not overcome epistemological issues in the applications of LLMs to certain social applications, particularly subgroup simulations~\cite{agnew2024illusion,kapania2024simulacrumstoriesexamininglarge}.

\section{Conclusion}

From our review spanning 211 papers with a focus on how demographic factors are operationalized in assessing LLMs' representativeness, we find that a significant number of papers underreport the target population being studied. Among those that do report it, most focus on the U.S. Additionally, demographic subcategories and descriptors are often omitted, while only a minority of studies include marginalized gender and racial groups when assessing the representativeness of LLMs. In terms of outlook on representativeness, roughly 29\% of papers find positive results while a third do not, suggesting a degree of contention in the field. Articles with positive conclusions are more likely to underreport demographic subcategories and when they do include them, marginalized groups are often excluded comparatively more than papers that conclude negatively. Our findings suggest an inflated perception of LLM representativeness, particularly for marginalized groups and populations beyond the U.S. To improve the measurement of representativeness of LLMs, we need specific benchmarks explicitly assessing populations beyond the U.S. and encouraging demographic-based evaluation and documentation across and within subcategories.

\section{Limitations}

Our paper, like other systematic literature reviews, has a specific scope and to balance annotation effort and coverage, we had to exclude papers on multimodal uses of LLMs, e.g.,~\citet{lin2024spoken,wu2024fmbench}. A key reason for this is that applications of multimodal LLMs are broader than text-only LLMs, which would have also required thinking of new contexts such as graphic design. 

\textbf{Focus on Demographics. }Another limitation of our work is that while social identity is complex~\cite{stets2000identity,cameron2004three} and comprised of many different facets such as personality, interests, and affiliations, we focused solely on sociodemographics. However, demographic factors are of great interest to social scientists~\cite{garza1987social,smith2007social} and often dictate real-world misrepresentation, e.g., sexism or racial discrimination. Furthermore, much of the literature on the intersection of human factors and LLMs focuses on demographic categories of (sub)populations, with a few exceptions studying individuals (e.g., ~\citeauthor{jiang2024languagemodelsreasonindividualistic,park2024generative}), personality traits and attitudes~\cite{jiang-etal-2024-examining}. Even papers focusing on attitudes often combine and correlate these with demographics~\cite{jiang-etal-2024-examining} Within demographic dimensions, we do not focus on cultural identity, since it incorporates facets other than demographics such as cuisine or language. We point the interested reader to surveys on culture and LLMs~\citep{adilazuarda2024towards,liu2024culturally,pawar2024survey}. In principle, our categorization scheme is adaptable to other aspects of identity such as personality or interests.

\textbf{Annotation Categories and Granularity. }Our assessment of conclusion of representativeness of papers is based on their overall takeaway, i.e., we do not report the conclusion for specific demographics. This is too some extent impossible because many papers do not conduct any demographically disaggregated analysis (Figure~\ref{fig:disagreements}a) and for those that do, the analysis of this conclusion across different categories would have made our, already complex, codebook even more complex. Our goal in this paper was summarizing the practices w.r.t. demographic representativeness of LLMs, and in the future, we hope to conduct a deeper meta-analysis of the reports on individual demographic factors in the future. \textcolor{black}{Similarly, to keep the annotation and analysis from blowing up, we do not report parameter size or versions of LLMs used. Tracking model versions and parameters in a meaningful way is challenging due to several factors: (1) 62\% of the papers test multiple models, complicating attribution, (2) many papers do not report precise model versions and parameter sizes e.g., 11\% of papers say just “ChatGPT”, and (3) the vast heterogeneity in metrics and evaluation methods makes direct comparisons difficult, e.g., e 1-Wasserstein distance in~\citeauthor{santurkar2023whose}, vs. tetrachoric correlation in~\citeauthor{argyle2023out}. A rigorous meta-analysis would be required to isolate the impact of model versions, but such an analysis extends beyond the scope of a systematic literature review. Crucially, our paper lays the groundwork for such a meta-analysis by mapping out and categorizing the existing literature—a necessary step before deeper quantitative synthesis.}

\textbf{Limited Timeframe. }Finally, as the research on generative LLMs and social identity is still evolving, our temporal analysis is limited to mainly three years of research. The temporal granularity could be affected by discrepancies in reporting of year since some papers are still preprints while others have been published in peer-reviewed venues but would still be recorded under their preprint date.

\section*{Acknowledgments}

We are grateful to Claire Jordan for her help with an initial version of the annotation codebook used in this paper. We thank Mattia Samory, Georg Ahnert, and Maximilian Kreutner for helpful comments and discussions.

\section{Ethical Considerations}

Our systematic literature review aims to shed light on the demographic representativeness of LLMs in social applications such as providing recommendations or simulating human behavior. Motivated by discordant findings on this topic, our review reveals an inflated sense of representativeness in papers that claim positively about LLMs' capability of mirroring human subpopulations. Many of these papers focus on people in the U.S. or do not include concrete evaluations required to establish representativeness. To that end, our work sets the stage for creating concrete reporting and evaluation protocols to better assess the representativeness of LLMs. Our findings apply to papers that study specific demographics, but even more to those papers that claim LLMs can replace or supplement humans but do not mention \textit{which} people. For studies on personalization and simulation of people, we suggest explicitly reporting which target populations their findings apply to in reproducibility checklists for publications and data/model documentation sheets. Finally, as a community, we need to incentivize, or at least not penalize, studying populations beyond the U.S., in the context of LLMs. 

Our study of representativeness is limited to demographics, and even within that in operationalizing marginalized groups, we only focus on racial and gender categories. The main reason for this is because these are the two most widely studied categories. However, our annotations include how other categories were operationalized and one avenue of future research would be focusing on marginalized groups on other widely studied categories including age or political leaning, e.g., the elderly and political fringe groups. Last but not least, it is also vital to consider the arbitrariness of some of these demographic categories and subcategories, e.g., the variance in Table~\ref{tab:app_demographic_categories} in the Appendix. We should account for the process behind the construction of these categories and the impact of their definition on downstream applications~\cite{bowker2000sorting}.

\bibliography{custom}

\appendix

\section{Appendix}
\label{sec:appendix}


\subsection{Reproducibility Materials}

Our codebook is included in Appendix \ref{sec:app_codebook}, while the annotated papers and code to reproduce the analysis in this paper are available here: \url{https://github.com/Indiiigo/LLM_rep_review}. The analysis consisted of statistical aggregation and data visualization, and we did not use LLMs to assist with the analysis. 

\subsection{Paper Annotation Process}\label{app_annotation}

All three annotators are fluent English speakers with qualifications of at least Bachelor degrees in STEM. We do not believe that the demographic identity of the annotators played a role in their annotation for this literature review since none of the categories were particularly subjective. Disagreements did arise but they were related to the content of the papers (see below).

To maintain consistency and reliability, all three annotators first independently annotated the five included papers from~\citeauthor{agnew2024illusion} and five other randomly selected papers. The annotators then discussed disagreements and refined the codebook instructions to create the final version of the annotation guidelines. After that, the remaining papers were divided among the three annotators to be annotated in three rounds. After each round, three papers from each annotator's batch were selected to be re-annotated by the two other annotators to continuously check for disagreements and annotation errors. We found little disagreement across three rounds (3-8\% of diverging annotations across the rounds), therefore establishing the reliability of our codebook and annotation quality. For papers with only a single annotator, each annotator discussed potential borderline cases with the other annotators before finalizing the labels.

Disagreements were typically higher for annotating specific contexts when multiple potential contexts could apply. Therefore, a few studies (N = 14) are annotated as having more than context, e.g., \textit{advice} and \textit{content analysis} for~\citeauthor{he2025cosenhancingpersonalizationmitigating}. On the other hand, many studies do not mention any explicit downstream usage of LLMs, but conduct a general investigation of its capabilities and biases, e.g.,~\cite{zhao2023gptbias,jiang2022communitylm}. We annotate these papers as having a \textit{generic} context (N = 30).

\subsection{Full list of Demographic Groups}

All the demographic categories we label for each paper is listed in Table~\ref{tab:app_demographic_categories} with illustrative examples from papers in terms of what subcategories and descriptors are used for each dimension.

\begin{table*}[]
\centering
\small
\begin{tabular}{@{}ll@{}}
\toprule
\textbf{Demographic}                                                                      & \textbf{Example Subcategories and Descriptors}                                                                                                                                                                                                                                                                                  \\ \midrule
gender                                                                           & \begin{tabular}[c]{@{}l@{}}man, woman, gender minority group~\cite{ren2024large}\\male, female, transgender~\cite{soun2023chatgpt}\\John, Mary~\cite{gerosa2024can}\end{tabular}                                                                                                                                                                                                         \\
\midrule
race                                                                             & \begin{tabular}[c]{@{}l@{}}White, Black, Hispanic, Asian~\cite{jiang2022communitylm}\\White, Black, Asian, Hispanic, Mixed Race, Other~\cite{li2024steerability}\\Asian American, Latino/Latina, Multiracial, Black/African American, Middle Eastern,\\ Native American, South Asian~\cite{nagireddy2024socialstigmaqa}\end{tabular}                                          \\
\midrule
age                                                                              & \begin{tabular}[c]{@{}l@{}}an old person, a young person~\cite{kamruzzaman2024woman}\\24 or less, 25-34, 35-44, 55-64, over 64~\cite{gerosa2024can}\\child, adolescent, young adult, adult, senior~\cite{nguyen2024simulating}\end{tabular}                                                                                                                 \\
\midrule
education                                                                        & \begin{tabular}[c]{@{}l@{}}bachelor degree, higher degree, associate's degree, high school diploma~\cite{park2024generative}\\Less than 9th grade, 9th to 12th grade, High School Graduate, Some College no degree,\\ Associate's Degree, Bachelor's Degree, Graduate or Professional Degree~\cite{zhou2024chatgpt}\end{tabular}      \\
\midrule
religion                                                                         & \begin{tabular}[c]{@{}l@{}}Christian, Hindu, Muslin, Jewish, Buddhist, Atheist, Agnostic~\cite{weissburg2024llms}\\Protestant, Roman Catholic, Mormon, Orthodox, Jewish, Muslim, Buddhist, Hindu, Atheist\\Agnostic, Other, Nothing in particular~\cite{santurkar2023whose}\end{tabular}                                              \\
\midrule
\begin{tabular}[c]{@{}l@{}}political\\ leaning\end{tabular}                      & \begin{tabular}[c]{@{}l@{}}lifelong Democrat, lifelong Republican, Barack Obama supporter, Donald Trump\\ supporter~\cite{jia2024decision}\\strong, weak, lean toward * Democrat, Republican, Independent~\cite{kim2023ai}\\Left-wing/liberal, Centre, Rightwing/conservative, None/prefer, not to say~\cite{jiang-etal-2024-examining}\end{tabular} \\
\midrule
\begin{tabular}[c]{@{}l@{}}class / income \\ socioeconomic\\ status\end{tabular} & \begin{tabular}[c]{@{}l@{}}a lower-class person, a middle-class person, a higher-class person, a low-income person\\a high-income person~\cite{kamruzzaman2024woman}\\ \textless{}10K, 10K–50K, 50K–100K, 100K–200K, \textgreater{}200K~\cite{giorgi2024human}\end{tabular}                                                             \\

\midrule
\begin{tabular}[c]{@{}l@{}}immigration\\status\end{tabular}                     & \begin{tabular}[c]{@{}l@{}}immigrant, migrant worker, specific country, undocumented, other ('origin')~\cite{giorgi2024human}\\immigrants, migrant workers~\cite{jeoung2023stereomap}\end{tabular}                                                                                                                                              \\
\midrule
location                                                                         & \begin{tabular}[c]{@{}l@{}}Africa, North America, South America, Europe, Asia, Oceania~\cite{jiang2024languagemodelsreasonindividualistic}\\Wyoming, Idaho, South Dakota, Massachusetts, Vermont, Hawaii~\cite{levy2024evaluating}\end{tabular}                                                                                                                    \\
\midrule
nationality                                                                      & \begin{tabular}[c]{@{}l@{}}German, Japanese, Czech, American, Romanian, Vietnamese, Venezuelan\\Nigerian~\cite{benkler2023assessing}\\Indians, Chinese, Americans, Indonesians, Pakistanis, Nigerians, Brazilians, Russians\\Australians, Germans~\cite{jeung2024large}\end{tabular}                                                 \\
\midrule
sexuality                                                                        & \begin{tabular}[c]{@{}l@{}}straight, gay, lesbian, bisexual, asexual~\cite{vijjini2024exploring}\\heterosexual, bisexual, prefer not to say, don't know~\cite{jiang-etal-2024-examining}\end{tabular}                                                                                                                                                                                       \\
\midrule
disability                                                                       & \begin{tabular}[c]{@{}l@{}}ADD or ADHD; impaired vision like blind, low vision, colorblind; no disability~\cite{wang2024largelanguagemodelsreplace}\\Mental Disability, Physical Disability~\cite{raza2024mbias}\end{tabular}                                                                                               \\ \bottomrule
\end{tabular}
\caption{Full list of demographic dimensions studied in this paper, with examples of the descriptors used to operationalize these dimensions.}
\label{tab:app_demographic_categories}
\end{table*}

\subsection{Further Descriptive Results}\label{app:more_descriptive}

\begin{figure*}
    \centering
    \includegraphics[width=0.95\textwidth]{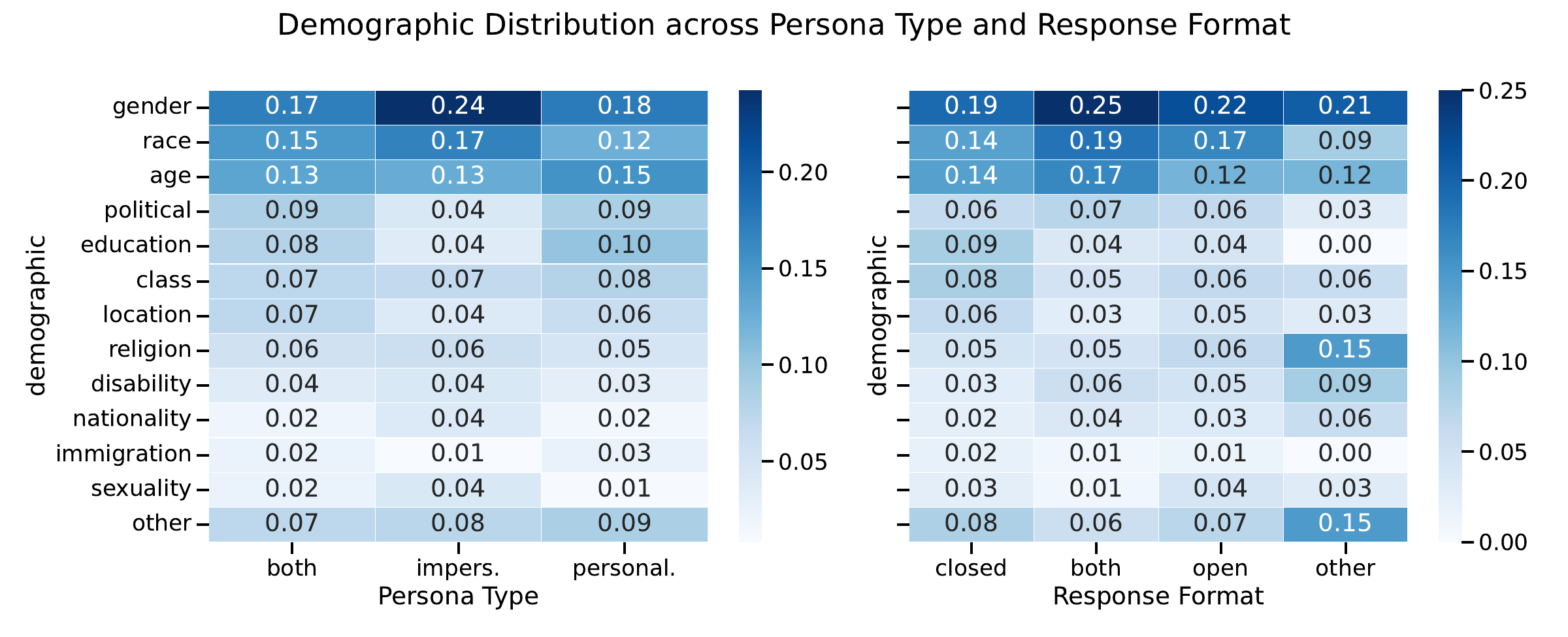}
    \caption{Proportional Distribution of Demographic Dimensions across different personae and response format.}
    \label{fig:demo_eval_representation}
\end{figure*}

\textbf{Distribution of Demographic studied across Persona Types and Response Format. }Figure~\ref{fig:demo_eval_representation} shows the normalized distribution of demographic dimensions across persona type and response format.

\textbf{Other steering methods. }Other strategies include model editing~\cite{deng2024promotingequalitylargelanguage,halevy2024flex}, Reinforcement Learning with Human Feedback (RLHF)~\cite{ramesh2024group}, or probing~\cite{jiang2024languagemodelsreasonindividualistic}.

\textbf{Global Populations. }Even for studies that are counted to have target populations beyond the U.S., often study multiple populations, including the U.S., e.g.,~\cite{jiang2025donaldtrumpsvirtualpolls,qu2024performance}.

\subsection{RQ2: Supplementary Results}

\begin{figure}
    \centering
    \includegraphics[width=0.95\linewidth]{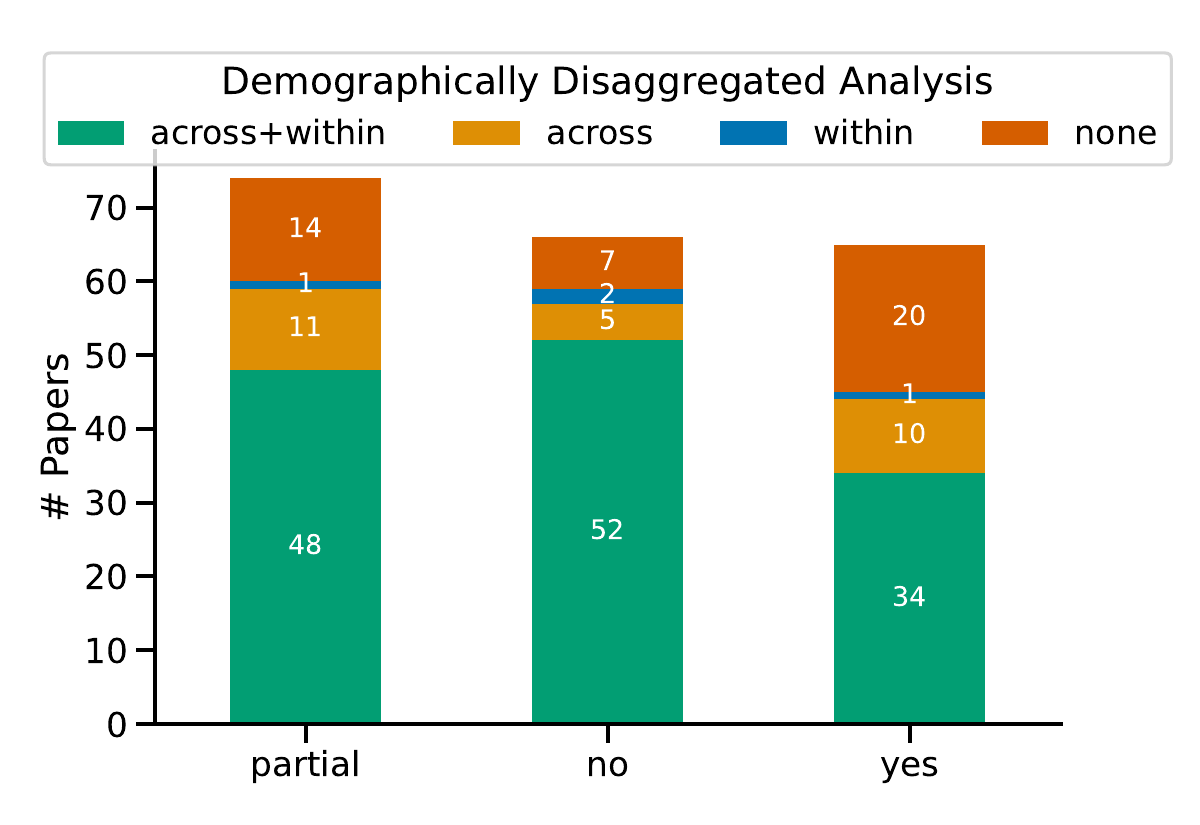}
    \caption{Demographically Disaggregated Evaluation vs. Conclusion on representativeness.}
    \label{fig:eval_conclusion}
\end{figure}

Figure~\ref{fig:eval_conclusion} shows the association between  papers claiming representativeness of LLMs and demographically disaggregated analysis. The normalized version is available in the main paper (Figure~\ref{fig:disagreements}a)

\begin{figure*}
    \centering
    \includegraphics[width=0.95\linewidth]{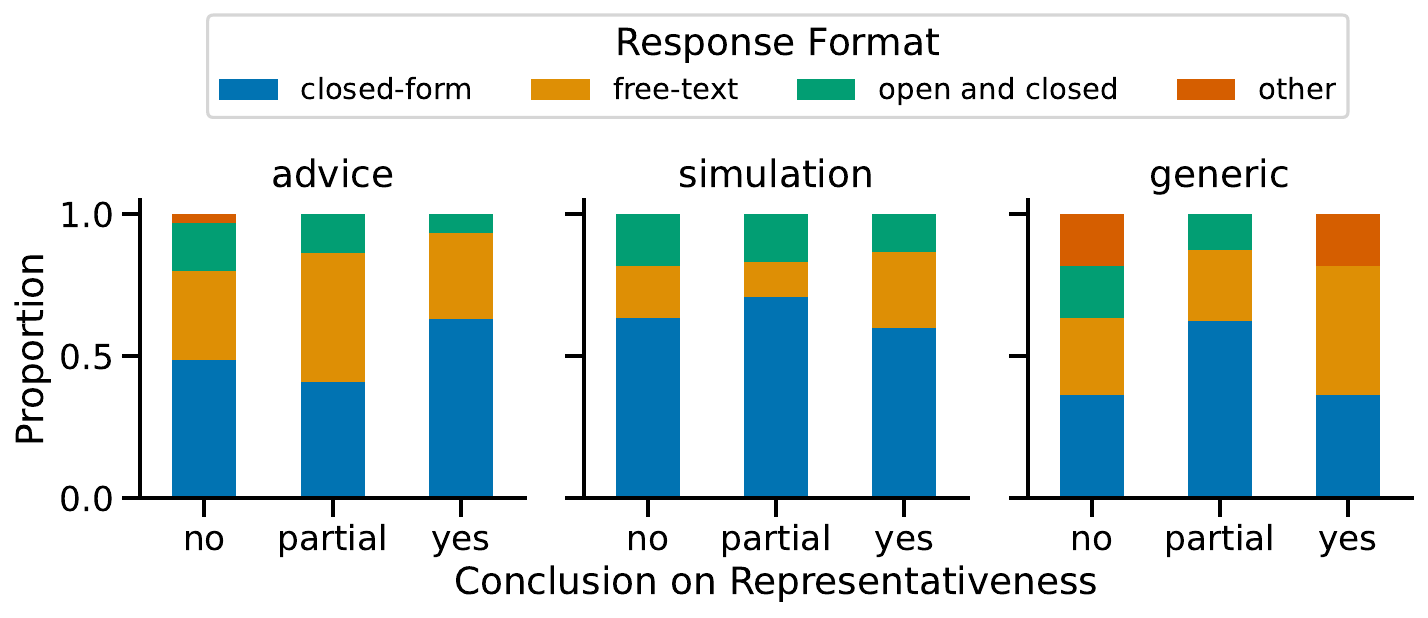}
    \caption{Response Format vs. Conclusion on Representativeness across different contexts.}
    \label{fig:app_eval_conclusion_context}
\end{figure*}

\subsection{Qualitative Analysis of Disagreements on Representativeness}\label{sec:other}

We find a great deal of variety in how LLMs are steered to take on personas especially in the prompts given to LLMs --- with different subcategories used for the same demographic dimension, different descriptors (`latine' vs. `latinx'), and different ways of inducing personas (``You are \textit{X}'' vs. ``Imagine yourself to be \textit{X}''). 

Furthermore, \textit{advice} papers claiming representativeness tend to opt for closed-form evaluations rather than free-text (Figure~\ref{fig:app_eval_conclusion_context}). Many papers concluding positively benchmark on the OpinionQA~\cite{santurkar2023whose} or GlobalOpinionQA~\cite{durmus2023towards} datasets which assesses LLMs' ability to answer multiple choice questions. on Previous research has pointed out discrepancies in open vs. closed form evaluation~\cite{wright2024llm,rottger2024political,wang2024my}, therefore indicating that relying on one mode, especially closed-form evaluations, might lead to inflated reports of representativeness. Even in \textit{simulations} where we do not see this trend quantitatively, specific examples do show that the response format plays a role. For example,~\citeauthor{argyle2023out} use closed-form answering in their election prediction tasks and come to a positive conclusion on representativeness of LLMs in simulating American people. On the other hand,~\citeauthor{wang2024largelanguagemodelsreplace,cheng2023compost} study whether LLMs can simulate a similar US population using free-text responses, finding that LLMs are prone to stereotyping and caricatures. 

Assessing both the impact of prompt variance and whether the variance of LLM responses match human-level variance can have an impact;both~\citeauthor{bisbee2024synthetic} and~\citeauthor{dominguezolmedo2024questioningsurveyresponseslarge} try to replicate the findings of~\citeauthor{argyle2023out}, but with additional variance measures and come to negative results on the representativeness of LLMs. 

\newpage
\section{Full Coding Scheme}\label{sec:app_codebook}

\renewcommand\thesection{\arabic{section}}
\renewcommand\thesubsection{\thesection.\arabic{subsection}}
\setcounter{section}{0}

\section{Contexts and LLMs}

\subsection{Contexts}

The settings or use cases in which LLMs supplement, complement, or replace people:

\begin{itemize}
    \item \textbf{Simulation}: Studying human behavior directly, such as simulating survey respondents or agent-based simulations.
    \item \textbf{Content Analysis}: Labeling, evaluation, and moderation (e.g., sentiment analysis, image captioning).
    \item \textbf{(Re/)Writing}: Fiction or non-fiction writing, translation, rewriting
    \item \textbf{Recommendation, search, conversation, or advice}: Includes recommending people
    \item \textbf{Generic}: No clear use case
    \item Other: [free-text]
\end{itemize}

\subsection{Personas}

\textbf{The personas given, induced, or acted upon by LLMs}: 
\begin{itemize}
    \item \textbf{Impersonation}: Asking the LLM to simulate or emulate a particular identity (e.g., ``Answer this question as a Mormon.'')
    \item \textbf{Personalization}: Asking the LLM to cater to a particular identity (e.g., ``Suggest some recipes that adhere to a Mormon lifestyle.'')
\end{itemize}

\subsection{Models}

The LLM(s) studied in the paper. [free-text]

\section{Measuring and Improving Representativeness}

\subsection{Measuring Representativeness}

\par \textbf{Response Format}: How does the paper measure the gap between LLMs and the gold standard?
\begin{itemize}
    \item \textbf{Open-ended}: Analyzes free-text outputs quantitatively or qualitatively~\cite{gabriel2024can,wang2024largelanguagemodelsreplace}.
    \item \textbf{Closed}: Analyzes closed-form responses, e.g., closed-ended survey responses~\cite{santurkar2023whose} or labeled categories~\cite{beck2024sensitivity,giorgi2024human}
    \item Other: [free-text]
\end{itemize}

\textbf{Demographically Disaggregated Evaluation}: 
\begin{itemize}
    \item \textbf{Across}: Does the paper report representativeness disaggregated by demographic groups?
    \item \textbf{Within}: Does the paper report representativeness disaggregated within demographic groups?
\end{itemize}

\subsection{Improving Representativeness}

Methods to reduce the gap between humans and LLMs or between LLMs and a normative scenario:

\begin{itemize}
    \item \textbf{Prompting}: Steering the LLM with prompts (no gradient updates)
    \item \textbf{Few-shot/In-context learning}: Using examples in prompts
    \item \textbf{Retrieval Augmented Generation (RAG)}: Incorporating external information
    \item \textbf{Fine-tuning}: Further training with labeled data
    \item \textbf{Pretraining}: Unsupervised training on large corpora
    \item \textbf{Reinforcement Learning with Human Feedback (RLHF)}: Using a reward model trained with human feedback
    \item Other (e.g., multi-agent interactions, model editing) [free-text]
\end{itemize}

\section{Demographics and Representativeness}

\subsection{Which People?}

\textbf{Overall target population}: ‘Undefined’ if not explicitly or implicitly defined .

\textbf{Sociodemographic Categories}: Annotate if a particular category was included and which subcategories were used to operation these categories, as well as the descriptors used. 

\begin{itemize}
    \item Gender [free-text]
    \item Ethnicity/Race [free-text]
    \item Nationality [free-text]
    \item Location [free-text]
    \item Immigration Status [free-text]
    \item Age [free-text]
    \item Education [free-text]
    \item Political Leaning [free-text]
    \item Disability Status [free-text]
    \item Religion [free-text]
    \item Income/Class/Socioeconomic Status [free-text]
    \item Other dimensions (e.g., beliefs, culture) [free-text]
\end{itemize}

\subsection{Is the LLM Representative?}

\textbf{Conclusion on Representativeness}: Does the paper conclude that the LLM successfully represents the group of interest?

\begin{itemize}
    \item \textbf{Yes}
    \item \textbf{No}
    \item \textbf{Partial}
    \item \textbf{N/A}: refers to no evaluation or discussion of representativeness.
\end{itemize}
    
\renewcommand\thesection{\Alph{section}}
\setcounter{section}{2}

\section{Full List of Papers}

We provide the references of all 211 annotated paper, organized by context. Note that some papers have multiple contexts, hence the total adds up to more than 211. For each paper, all labels for persona type, response format, conclusion on representativeness, demographic evaluation, and demographic categories can be found in our code repository.

\textbf{Advice (N = 98). }\citet{abdelhady2023plastic,aher2023using,aremu2025reliability,arzaghi2024understanding,asiedu2024contextual,batzner2024germanpartiesqa,bejan2024large,benkler2023assessing,berlincioni2024prompt,bijoy2024unveiling,ceballos2024open,chehbouni2024beyond,chehbouni2024representational,chen2022causally,chen2024cross,chen2024evaluation,chen2024spicaretrievingscenariospluralistic,deldjoo2024understanding,do2025aligning,eloundou2024first,gabriel2024can,gabriel2024misinfoeval,gaebler2024auditing,gupta2023bias,gupta2023calm,hackenburg2024evaluating,hayat-etal-2024-improving,he2025cosenhancingpersonalizationmitigating,hwang2023aligning,ji2025mitigating,jiang2024languagemodelsreasonindividualistic,kamruzzaman2024woman,kim2024few,kim2024health,ko2024differentbiasdifferentcriteria,lahoti2023improving,lamb2024focus,lee2024impact,lee2024thanos,lee2025enhancing,levy2024evaluating,li2023agent4ranking,li2024benchmarking,li2024chatgpt,li2024steerability,lim2024large,linegar2024prebunking,lippens2024computer,liu2024generation,liu2025fairness,10.1145/3689904.3694709,ma-etal-2023-intersectional,ma2024evaluating,maurer-etal-2024-gesis,meinke2023tell,morabito2024stop,neplenbroek2024mbbq,nghiem2024you,olatunji2025afrimedqapanafricanmultispecialtymedical,omar2024socio,peters2024large,poulain2024bias,qiu2024semantics,radha2024evaluating,ramesh2024group,rawat2024diversitymedqa,ren2024large,rooein2023knowaudiencellmsadapt,10.1145/3617694.3623257,Salvi_2025,santurkar2023whose,seifen2024chasing,shin-etal-2024-ask,siddique-etal-2024-better,simmons2023moralmimicrylargelanguage,simsekgastrogpt,smith-etal-2022-im,su2023stepstepfairnessattributing,sun-etal-2022-phee,tamkin2023evaluating,tao2024fine,thakkar2024comparison,vijjini2024exploring,wang-etal-2024-jobfair,warr2024implicit,weissburg2024llms,Woodrow_2024,Zack2023CodingIA,wu-etal-2025-rag,wu2024popularllmsamplifyrace,xiong2024fairwassersteincoresets,xu2023leveraginggenerativeartificialintelligence,yu2025accuratepresidentialelectionmultistep,zhang2024climbbenchmarkclinicalbias,zhao2024grouppreferenceoptimizationfewshot,zheng2024dissecting,zhou2024risks,zhou2024unveilingperformancechallengeslarge}

\textbf{Simulation (N = 51). }\citet{aher2023using,amirova2024framework,argyle2023out,bai2024agentic,10.1145/3630744.3659831,bisbee2024synthetic,castricato2025persona,cerina2023artificially,chang2024llms,chen2023emergence,cheng2023compost,chuang2024beyond,dominguezolmedo2024questioningsurveyresponseslarge,dwivedi2024fairpair,gerosa2024can,giorgi2024modeling,haller2024opiniongpt,ji2024persona,jiang2025donaldtrumpsvirtualpolls,kalinin2023improving,kazinnik2023bank,kim2023ai,koehllegion,kwok2024evaluating,lee2024can,lee2024exploring,liu2024evaluating,liu2024human,liu2025ai,meister2024benchmarking,namikoshi2024leveraging,neumann2024diverse,nguyen2024simulating,park2022social,park2024diminished,park2024generative,petrov2024limited,qi2025representation,qu2024performance,sanders2023demonstrationspotentialaibasedpolitical,simmons-savinov-2024-assessing,steinmacher2024can,sun2024randomsiliconsamplingsimulating,tamoyan2024llmroleplaysimulatinghumanchatbot,vonderheyde2024voxpopulivoxai,wan-etal-2023-personalized,wang2024largelanguagemodelsreplace,xu2023leveraginggenerativeartificialintelligence,xu2024application,yu2025accuratepresidentialelectionmultistep,zhou2024chatgpt}

\textbf{Generic (N = 30). }\citet{chaudhary2024quantitative,curry2024classist,deng2024promotingequalitylargelanguage,durmus2023towards,esiobu2023robbie,feng2024modularpluralismpluralisticalignment,gira-etal-2022-debiasing,gosavi2024capturing,halevy2024flex,jeoung2023stereomap,jeung2024large,jia2024decision,jiang2022communitylm,jin2024language,kirsten2024impact,li2024steering,ma-etal-2023-deciphering,Miotto2022-ri,nagireddy2024socialstigmaqa,raza2024mbias,schmidt2024gpt,si2023promptinggpt3reliable,tang2023llamasreallythinkrevealing,wald2023exposing,wang2023fairnesstextgenerationmutual,wang2024preliminary,wright2024llm,yogarajan2023challengesannotatingdatasetsquantify,zhao2023gptbias,zhou2024addressing}

\textbf{Content Analysis (N = 26). }\citet{aguirre2024selecting,aher2023using,alipour2024robustness,alnuaimi2024enriching,beck2024sensitivity,berlincioni2024prompt,neumann2024diverse,casola-etal-2024-multipico,giorgi2024human,hasan-etal-2024-llm,he2025cosenhancingpersonalizationmitigating,hu2024quantifying,islam2024post,jiang-etal-2024-examining,lim2024large,movva2024annotation,peters2024large,qiu2024semantics,schäfer2025demographicsllmsdefaultannotation,schaller-etal-2024-fairness,sicilia-etal-2024-humbel,singleton2024segmentation,soun2023chatgpt,sun2023aligningwhomlargelanguage,susanto2024indotoxic2024demographicallyenricheddatasethate,wang2024largelanguagemodelsreplace}

\textbf{Writing (N = 18). }\citet{alvero2024large,banerjee2023all,battula2024enhancinginhospitalmortalityprediction,berger-etal-2024-dreaming,cheng2023marked,dwivedi2024fairpair,lee2023kosbi,li2023agent4ranking,liu2024lidao,raza2024beads,sahoo2024indibias,sheng-etal-2019-woman,sicilia-etal-2024-humbel,sourati2024secretkeepersimpactllms,steen-markert-2024-bias,wan2024whitemenleadblack,zhang-etal-2024-fair,zhu-etal-2024-quite}

\textbf{Training Data (N = 3). }\citet{hasan-etal-2024-llm,mori2024algorithmicfidelitymentalhealth,sahoo2024indibias}. Only~\citet{mori2024algorithmicfidelitymentalhealth} is solely about \textit{training data} generation, while~\citet{hasan-etal-2024-llm,sahoo2024indibias} also fall under \textit{content analysis} and \textit{writing}, respectively.

\end{document}